\documentstyle[12pt]{article}
\catcode`\@=11
\def\section{\@startsection {section}{1}{0pt}{-3.5ex plus -1ex minus
 -.2ex}{2.3ex plus .2ex}{\raggedright\large\bf}}
\catcode`\@=12
\newskip\humongous \humongous=0pt plus 1000pt minus 1000pt

\newif\ifdtup

\def\oldreffmt#1{\rlap{[#1]} \hbox to 2\parindent{}}

\def\figfmt#1{\rlap{Figure {#1}} \hbox to 1in{}}

\def\bra#1{\left\langle #1\right|}
\def\ket#1{\left| #1\right\rangle}

\def\ltap{\raisebox{-.4ex}{\rlap{$\sim$}} \raisebox{.4ex}{$<$}}
\def\gtap{\raisebox{-.4ex}{\rlap{$\sim$}} \raisebox{.4ex}{$>$}}

\def\beq{\begin{equation}}
\def\eeq{\end{equation}}
\def\bea{\begin{eqnarray}}

\def\com#1#2{
        \left[#1, #2\right]}
\def\eea{\end{eqnarray}}
\def\ap#1,#2,#3#4{           {\it Ann. Phys. (NY) }{\bf #1}: #2 (19#3#4)}
\def\apj#1,#2,#3#4{          {\it Astrophys. J. }{\bf #1}: #2 (19#3#4)}
\def\apjl#1,#2,#3#4{         {\it Astrophys. J. Lett. }{\bf #1}: #2 (19#3#4)}
\def\app#1,#2,#3#4{          {\it Acta Phys. Polon. }{\bf #1}: #2 (19#3#4)}
\def\com#1,#2,#3#4{          {\it Comm. Math. Phys. }{\bf #1}: #2 (19#3#4)}
\def\ib#1,#2,#3#4{           {\it ibid. }{\bf #1}: #2 (19#3#4)}
\def\nat#1,#2,#3#4{          {\it Nature (London) }{\bf #1}: #2 (19#3#4)}
\def\np#1,#2,#3#4{           {\it Nucl. Phys. }{\bf B#1}: #2 (19#3#4)}
\def\pl#1,#2,#3#4{           {\it Phys. Lett. }{\bf #1B}: #2 (19#3#4)}
\def\pr#1,#2,#3#4{           {\it Phys. Rev. }{\bf D#1}: #2 (19#3#4)}
\def\prep#1,#2,#3#4{         {\it Phys. Rep. }{\bf #1}: #2 (19#3#4)}
\def\prl#1,#2,#3#4{          {\it Phys. Rev. Lett. }{\bf #1}: #2 (19#3#4)}
\def\pro#1,#2,#3#4{          {\it Prog. Theor. Phys. }{\bf #1}: #2 (19#3#4)}
\def\rmp#1,#2,#3#4{          {\it Rev. Mod. Phys. }{\bf #1}: #2 (19#3#4)}
\def\sp#1,#2,#3#4{           {\it Sov. Phys.-Usp.}{\bf #1}: #2 (19#3#4)}

\def\zp#1,#2,#3#4{           {\it Zeit. fur Physik }{\bf #1}: #2 (19#3#4)}
\catcode`\@=11
\def\eqnarray{\stepcounter{equation}\let\@currentlabel=\theequation
\global\@eqnswtrue
\global\@eqcnt\z@\tabskip\@centering\let\\=\@eqncr
\gdef\@@fix{}\def\eqno##1{\gdef\@@fix{##1}}%
$$\halign to \displaywidth\bgroup\@eqnsel\hskip\@centering
  $\displaystyle\tabskip\z@{##}$&\global\@eqcnt\@ne
  \hskip 2\arraycolsep \hfil${##}$\hfil
  &\global\@eqcnt\tw@ \hskip 2\arraycolsep $\displaystyle\tabskip\z@{##}$\hfil
   \tabskip\@centering&\llap{##}\tabskip\z@\cr}

\def\@@eqncr{\let\@tempa\relax
    \ifcase\@eqcnt \def\@tempa{& & &}\or \def\@tempa{& &}
      \else \def\@tempa{&}\fi
     \@tempa \if@eqnsw\@eqnnum\stepcounter{equation}\else\@@fix\gdef\@@fix{}\fi
     \global\@eqnswtrue\global\@eqcnt\z@\cr}

\newcount\hour
\newcount\minute
\newtoks\amorpm
\hour=\time\divide\hour by60
\minute=\time{\multiply\hour by60 \global\advance\minute by-\hour}
\edef\standardtime{{\ifnum\hour<12 \global\amorpm={am}%
	\else\global\amorpm={pm}\advance\hour by-12 \fi
	\ifnum\hour=0 \hour=12 \fi
	\number\hour:\ifnum\minute<10 0\fi\number\minute\the\amorpm}}
\edef\militarytime{\number\hour:\ifnum\minute<10 0\fi\number\minute}

\def\draftlabel#1{{\@bsphack\if@filesw {\let\thepage\relax
   \xdef\@gtempa{\write\@auxout{\string
      \newlabel{#1}{{\@currentlabel}{\thepage}}}}}\@gtempa
   \if@nobreak \ifvmode\nobreak\fi\fi\fi\@esphack}
        \gdef\@eqnlabel{#1}}
\def\@eqnlabel{}
\def\@vacuum{}
\def\marginnote#1{}
\def\draftmarginnote#1{\marginpar{\raggedright\scriptsize\tt#1}}
\def\draft{
	\pagestyle{plain}
	\overfullrule=2pt
        \oddsidemargin -.5truein
        \def\@oddhead{\sl \phantom{\today\quad\militarytime} \hfil
        \smash{\Large\sl DRAFT} \hfil \today\quad\militarytime}
        \let\@evenhead\@oddhead
        \let\label=\draftlabel
        \let\marginnote=\draftmarginnote
        \def\ps@empty{\let\@mkboth\@gobbletwo
        \def\@oddfoot{\hfil \smash{\Large\sl DRAFT} \hfil}
        \let\@evenfoot\@oddhead}
        \def\@eqnnum{(\theequation)\rlap{\kern\marginparsep\tt\@eqnlabel}%
        \global\let\@eqnlabel\@vacuum}  }

\catcode`\@=12

\def\lae{\smash{\,\lower .5 ex \hbox{$\,\stackrel<\sim\,$}}}
\def\gae{\smash{\,\lower .5 ex \hbox{$\,\stackrel>\sim\,$}}}

\def\L{{\cal L}}

\def\beq{\begin{equation}}
\def\eeq{\end{equation}}

\def\sutw{${\rm SU}(2)_W$}

\begin{document}
\begin{titlepage}
\begin{center}
April 5, 1993\hfill    WIS--93/26/March--PH

\vskip 1 cm

{\large \bf  New Bounds on Leptoquarks}

\vskip 1 cm

Miriam Leurer

\vskip 1 cm

{\em Department of Nuclear Physics\\
The Weizmann Institute\\
Rehovot 76100\\
ISRAEL}

\end{center}

\vskip 1 cm

\begin{abstract}

We derive new bounds on scalar leptoquark couplings from $K^0-\bar K^0$,
$D^0-\bar D^0$ and $B^0-\bar B^0$ mixing. Although leptoquarks contribute to
these processes only at one loop, their contribution is large, due to the lack
of GIM cancelation. Our bounds have two important features: (i) They bound
$g^4/M^2$, in contrast to the hitherto known bounds on $g^2/M^2$, and are
consequently stronger at high masses. (ii) The bound from $D^0-\bar D^0$ mixing
is the first FCNC bound in the up sector for chirally-coupled leptoquarks, and
is similar in strength to the $K^0-\bar K^0$ and $B^0-\bar B^0$ bounds.
Together, these bounds strongly constrain any leptoquark that couples to left
handed quarks.

\end{abstract}
\end{titlepage}
\newpage

In the last few years we have witnessed a renewed interest  in low-lying
leptoquarks. This has been stimulated on the one hand by the construction of
the $ep$ collider, HERA, which is an ideal machine for leptoquark searches, and
on the other hand by non-standard models which predict the existence of such
particles. There are already several bounds on leptoquarks masses and mixings
from existing $e^+e^-$ and $p\bar p$ machines \cite{Shanker}. These bounds can
be summarized as follows: (i) The most stringent bounds arise when a leptoquark
is allowed to couple to both left-handed (LH) and right-handed (RH) quarks.
The pseudo-scalar mesons, $\pi$, $K$ and $D$ can then decay leptonically
without the ``chiral suppression'' of the standard model. To avoid these
bounds, one usually demands that leptoquarks couple {\it chirally}, either to
left-handed or to right-handed quarks, but not to both. (ii) The strongest
bounds on {\it chirally} coupled leptoquarks arise from
flavour changing neutral current (FCNC) processes in the lepton and the down
sectors. It is therefore customary to demand that leptoquarks couple
``diagonally'' to these sectors, {\it i.e.} any leptoquark is allowed to couple
only
to a single lepton generation and a single down quark generation. (iii) The
bounds are derived by considering {\it tree-level}{} leptoquark contributions
to various processes. Therefore, once the masses are heavy relative to the
available energies, the bounds apply to $g^2/M^2$. At present, the highest
relevant energy is TRISTAN's with  $E\sim 60$ GeV, and ``heavy masses'' are
$M\gtap 200$ GeV.

In this paper we wish to point out a new set of bounds. These arise from the
one-loop contribution of the leptoquarks to neutral meson mixing. We first
present these bounds for general scalar leptoquarks, without any restriction to
``chiral'' or ``diagonal'' couplings: Consider the coupling of a scalar
leptoquark $\Phi$ to a particular lepton (or antilepton) $l$ and to the down
quarks $d$ and $s$:
\beq
\L  =  \left\{ \bar l \left[g^{d}_LP_L+g^d_RP_R\right] d
       +       \bar l \left[g^{s}_LP_L+g^s_RP_R\right] s\right\}\;,
\label{lagr}
\eeq
where the indices on the coupling constants indicate the flavour and the
chirality of the quark, and $P_L$ and $P_R$ are the LH and RH projection
operators. The interaction  (\ref{lagr}) leads to a new contributions to
$K^0-\bar K^0$ mixing via a loop of leptons and leptoquarks:
\beq
\Delta M_{12}=\frac{1}{192\pi^2M_{LQ}^2} \left|
(a_{L}^K)^2+(a_{R}^K)^2-(E_K+\frac{3}{2})a_{L}^Ka_{R}^K \right| f_K^2M_K
\;,
\label{mixK}
\eeq
where $a_{L}^K=g^d_L(g^s_L)^*$, $a_{R}^K=g^d_R(g^s_R)^*$ and $E_K$ is the
ratio of $\bra{\bar K^0}\bar d \gamma_5 s\ket{\vphantom{\bar K^0}0}
\bra{\vphantom{\bar K^0}0}\bar d \gamma_5 s\ket{\vphantom{\bar K^0}K^0}$
to
$\bra{\bar K^0}\bar d \gamma^\mu\gamma_5 s\ket{\vphantom{\bar K^0}0}
\bra{\vphantom{\bar K^0}0}\bar d
\gamma_\mu\gamma_5 s\ket{\vphantom{\bar K^0}K^0}$.
Demanding that the new contribution to $\Delta M_{12}$ does not exceed
the measured value we get:
\beq
\frac{1}{M_{LQ}^2} \left|(a_{L}^K)^2 + (a_{R}^K)^2 + (E_K+\frac{3}{2}) \,
a_{L}^K  a_{R}^K \right|
\, \leq \, 5.2\cdot 10^{-10}~~{\rm GeV^{-2}}\; .
\label{boundK}
\eeq
There are a few important points we would like to impress on the reader: First,
the contribution (\ref{mixK}) to $K^0-\bar K^0$ mixing is particularly large,
despite being one-loop. This is because there is no Glashow-Iliopoulos-Maiani
(GIM) mechanism for scalar leptoquarks and
consequently no analogue to the standard suppression factor $m_c^2/M_W^2$. The
large contribution to the mixing translates, in equation (\ref{boundK}), to a
strong bound on the leptoquark coupling. Second, the contribution of the
lepton-leptoquark loop is independent of the nature of the lepton (or
antilepton). Corrections due to the lepton mass are small and were neglected.
Finally, if there are several contributions with various leptons and
leptoquarks running in the loop, the bound applies separately to each of them,
since there is no GIM mechanism here and no significant cancelations are
expected.

Repeating the same procedure for $D^0 - \bar D^0$ and $B^0 - \bar B^0$
mixing, we find:
\begin{eqnarray}
\frac{1}{M_{LQ}^2}
\left|(a_{L}^D)^2 + (a_{R}^D)^2 + (E_D+\frac{3}{2}) \, a_{L}^D a_{R}^D\right|
&\leq& 2.2\cdot 10^{-9}~~{\rm GeV^{-2}}\label{boundD}\\
\frac{1}{M_{LQ}^2}
\left|(a_{L}^B)^2 + (a_{R}^B)^2 + (E_B+\frac{3}{2})\, a_{L}^B a_{R}^B\right|
&\leq& 7.5\cdot 10^{-9}~~{\rm GeV^{-2}}\; ,
\label{boundB}
\end{eqnarray}
where $a^D_L=g^u_L(g^c_L)^*$, $a^D_R=g^u_R(g^c_R)^*$ and
$a^B_L=g^d_L(g^b_L)^*$, $a^B_R=g^d_R(g^b_R)^*$. Here we have used the FNAL--TPS
bound on $D^0-\bar D^0$ mixing \cite{FNAL}, which translates to $\Delta M\leq
1.5\cdot 10^{-4}$ eV at 95\% CL, and the PDG value
\cite{PDG} for $B^0 - \bar B^0$ mixing, which translates to $\Delta M\leq
5.0\cdot 10^{-4}$ eV at 95\% CL.

The bounds (\ref{boundK}--\ref{boundB}) have two important features: First,
they apply to $g^4/M^2$, in contrast to previous
bounds which apply to $g^2/M^2$. Our bounds therefore dominate at high
masses. As an example, consider bounds on FCNC in the down sector for chirally
coupled leptoquarks: Previously, the strictest bounds were derived from
$K\longrightarrow\pi\nu\bar\nu$ and $K_L\longrightarrow e^+e^-$ decays
(see table 1). Now we have also the $K-\bar K$ bound, which is more stringent
at masses of order 1 TeV.

The other important feature is the
$D^0-\bar D^0$ mixing bound. {This is the only {\it up} sector FCNC bound that
applies to chirally-coupled leptoquarks}, and it has significant consequences:
As mentioned above, all previous FCNC bounds for chirally-coupled
leptoquarks could be avoided by demanding that the leptoquarks couple
diagonally to the lepton and down sectors. But now we need to demand the
diagonality of the leptoquark couplings to the up sector as well. If the
leptoquark couples to RH quarks, all ``diagonality'' demands in the lepton,
down and up sectors can be satisfied. But, if the leptoquark couples to LH
quarks, Cabibbo-Kobayashi-Maskawa (CKM) mixing implies that we cannot
diagonalize its couplings to the down and the up sectors simultaneously. We
therefore find, in contrast to hitherto existing bounds, that FCNC bounds
cannot be completely evaded when a leptoquark couples to LH quarks.

To display the power of our new bounds we shall now discuss chiral leptoquarks
that couple to LH quarks, and ``as diagonally as possible'' to the first
generation. We will present the allowed regions in the coupling
constant--mass plane and see that large regions are excluded
when our bounds are
taken into account.

There are three types of scalar leptoquarks that couple to LH quarks: $S$, $D$
and $T$, which are a singlet, a doublet and a triplet of \sutw{} and carry
$Y= -1/3$, $-7/6$ and $-1/3$ respectively. In the following we will ignore
possible mass splitting within each of the leptoquark multiplets and the
possibility of mixing amongst the multiplets. The couplings to the fermions
are given by:
\begin{eqnarray}
\L_S &=& \sum_i \, \left(g_i~\bar e^c u^i_L + g'_i~\bar\nu^c d^i_L \right)
               \, S_{1/3}\nonumber\\
\L_D &=& \sum_i \, \left\{g_i~\bar e\, u^i_L D_{-5/3} + g'_i~\bar e\, d^i_L
	D_{-2/3} \right\} \nonumber\\
\L_T &=& \sum_i \, \left\{g_i~\bar \nu^c u^i_L T_{-2/3}
      + \frac{1}{\sqrt2} (g_i~\bar e^c u^i_L - g'_i~\bar\nu^c d^i_L) \, T_{1/3}
      + g_i~\bar e^c d^i_L T_{4/3} \right\}  \; ,
\label{inter}
\end{eqnarray}
where the $g_i$ and $g'_i$ couplings are related by:
\beq
g'_i=g_jV_{ji}\;,
\label{gCKM}
\eeq
with $V$ the CKM mixing matrix. The relation
(\ref{gCKM}) is the mathematical realization of the fact that we cannot fully
diagonalize the couplings: If, for example, we choose to diagonalize the
couplings to the down quark sector, demanding that $g'_2$ and $g'_3$ vanish,
then the couplings to the up sector cannot be completely diagonal: $g_2$ and
$g_3$ do not vanish, although they are suppressed by CKM factors. We will not
make an apriori decision as to which sector should be diagonalized, but only
demand that the $g_i$ ($i=2,3$) are CKM suppressed, that is:
\beq
g_i \, \ltap \, g_1V_{1i} \; .
\label{g2g3}
\eeq
Note that the same suppression
will then apply to the $g'_i$ couplings, $g'_i\,\ltap\, g'_1 V_{1i}$, and
that $g_1$ and $g'_1$ are equal up to a small correction, to be neglected here.

In the following we ignore the third generation since its couplings are
strongly suppressed, (by $V_{13}$). We are then reduced to a
two-generation picture, and parametrize our coupling constants as follows:
\begin{eqnarray}
g_1=g\cos\theta  ~~~~~ &~~&~~~~~
g'_1=g\cos(\theta + \theta_C) \nonumber\\
g_2=g\sin\theta
{}~~~~~ &{~~}& ~~~~~
g'_2=g\sin(\theta + \theta_C)
\label{defg}\;.
\end{eqnarray}
The $g$ and $\theta$ parameters defined in equation (\ref{defg}) have the
following interpretation: $g$ is the coupling constant of the leptoquark to LH
quarks, while $\theta$ determines the distribution of this coupling between the
first and second generation. Since we choose to couple the leptoquark mainly to
the first generation, $|\theta|$ is not larger than $\sim\theta_C$. The bounds
derived from experimental data apply to the $g_i$ and $g'_i$ couplings, and are
summarized in table 1.

We now translate the information of the table to bounds on $g^2$: The bound in
the first row applies directly to $g^2$ ($g$ and $g_1$ being equal up to a
small correction which we neglect) and is quadratic in the leptoquark mass.
This is the ``classical'' bound, which was known prior to this work.

The other bounds in the table are derived from FCNC processes. We rewrite
them as:
\begin{eqnarray}
2|g'_1g'_2|&=&g^2|\sin2(\theta+\theta_C)|\leq f(M)\label{dbound} \\
2|g_1g_2|&=&g^2|\sin2\theta|\leq g(M) \label{ubound} \;,
\end{eqnarray}
where $f$ and $g$ are functions of the leptoquark mass $M$. Equation
(\ref{dbound}) summarizes the FCNC bounds in the down sector: At low leptoquark
masses the rare $K$ decay bound dominates and $f(M)$ is quadratic in the
leptoquark mass, while in the high mass region the $K^0-\bar K^0$ bound
dominates and $f(M)$ is linear. Equation (\ref{ubound}) is the $D^0-\bar D^0$
bound, the only FCNC bound of the up sector. $g(M)$ is linear in the leptoquark
mass. One could evade the FCNC bounds in one of the sectors by choosing
$\theta=-\theta_C$ or $\theta=0$, but it is impossible to evade the bounds in
both sectors simultaneously.  Every $\theta$ will lead to a
bound on $g^2$. The weakest possible
bound corresponds to a compromise between the two
sectors, that is to $\theta(M)$ which is between $-\theta_C$ and 0:
\beq
\sin{2\theta(M)}=-\frac{\sin{2\theta_C}}{\sqrt{(\sin{2\theta_C})^2+
\left(\cos2\theta_C+f(M)/g(M)  \right)^2}}\;.
\label{theta}
\eeq
Then
the combined effect of the FCNC constraints in the
up and the down sectors
gives the bound
\beq
g^2\leq g^2_{max}=f(M)/\sin{2(\theta(M)+\theta_C)}=g(M)/\sin{2\theta(M)} \;.
\label{gmax}
\eeq
Note that in the high mass region $\theta(M)$ is
$M$-independent ($\sin{2\theta(M)}\approx-0.3$) and $g^2_{max}$ is linear in
$M$. The
new FCNC bound therefore becomes stronger than the classical quadratic bound at
high leptoquark masses, and it excludes new regions in the coupling
constant--mass plane. The classical bound and the new FCNC--combined bound are
shown in figures 1(a--c) for the various leptoquarks.

Summarizing, we have found new FCNC bounds on leptoquark couplings by examining
their one-loop contributions to neutral meson mixing. The new bounds are
particularly powerful for leptoquarks that couple to left handed quarks: In
this case, since the CKM mixing implies that FCNC cannot be avoided in both the
up and the down sectors, our $K^0-\bar K^0$ and $D^0-\bar D^0$ bounds combine
to bound the flavour {\it conserving} coupling constant. This bound is shown in
figures 1(a--c), and it excludes large regions in the coupling constant--mass
plane. Although these regions lie beyond HERA's kinematical limit, they are
significant for virtual leptoquark searches in HERA, and for searches in the
SSC and LHC.

\noindent
{\bf Acknowledgements}\\
I thank Neil Marcus for helpful conversations.

\newpage
\noindent {\large \bf Table Caption}

\noindent {\bf Table 1:} {
95\% CL experimental bounds on the $g_i$ and $g'_i$ coupling. Masses are in
GeV. Note that the first two bounds are quadratic in $M$, while the last two
are linear. The FCNC bounds are presented with an explicit $\sin\theta_C$
factor.  For $e^+e^-$ scattering the scale $\Lambda$ of nonstandard physics was
found in \cite{TOPAZ} to lie in the TeV range at 95\% CL. As a representative
number we take $\Lambda=5$ TeV. }

\vskip 1cm

\begin{center}
\begin{tabular}{|c|c|c|c|}\hline
& S & D & T\\ \hline
$|g_1|^2/M^2$ & $9.6\cdot 10^{-8}$
                        & $5.0\cdot 10^{-7}$
                        & $5.0\cdot 10^{-7}$ \\
                        & $\pi\longrightarrow e\nu $ \cite{Britton}
                        & $e^+e^-$\cite{TOPAZ}
                        & $e^+e^-$\cite{TOPAZ}\\ \hline
$|g'_1g'_2|/M^2$
                        & $4.5\cdot 10^{-8}\sin\theta_C$
                        & $1.2\cdot 10^{-7}\sin\theta_C$
                        & $8.9\cdot 10^{-8}\sin\theta_C$ \\
                        & $K\longrightarrow\pi\nu\bar\nu$ \cite{BNL}
                        & $K_L\longrightarrow ee$ \cite{KEK}
                        & $K\longrightarrow\pi\nu\bar\nu$ \cite{BNL} \\ \hline
$|g'_1g'_2|/M$ & $1.0\cdot 10^{-4}\sin\theta_C$
                        & $1.0\cdot 10^{-4}\sin\theta_C$
                        & $9.3\cdot 10^{-5}\sin\theta_C$\\
                        & $K^0-\bar K^0$ \cite{PDG}
                        & $K^0-\bar K^0$ \cite{PDG}
                        & $K^0-\bar K^0$ \cite{PDG} \\ \hline
$|g_1g_2|/M$   & $2.1\cdot 10^{-4}\sin\theta_C$
                        & $2.1\cdot 10^{-4}\sin\theta_C$
                        & $1.9\cdot 10^{-4}\sin\theta_C$\\
                        & $D^0-\bar D^0$ \cite{FNAL}
                        & $D^0-\bar D^0$ \cite{FNAL}
                        & $D^0-\bar D^0$ \cite{FNAL}    \\ \hline
\end{tabular}
\end{center}

\newpage
\noindent {\large \bf Figure Captions}

\noindent{\bf Figure 1(a--c):} Upper bounds on the leptoquark coupling
constant $g$ as a function of its mass $M$, when the leptoquark couples to LH
quarks of the first generation and to leptons of the first generation. Figures
(1a), (1b) and (1c) describe the bounds for the $S$, $D$ and $T$ leptoquark
respectively.  We show the direct and FCNC bounds, as well as the $4\pi$
perturbative upper limit.

\end{document}